# On-the-fly precision spectroscopy with a dual-modulated tunable diode laser and Hz-level referencing to a cavity

Shuangyou Zhang[1], Toby Bi[1,2], Pascal Del'Haye[1,2]*

[1]Max Planck Institute for the Science of Light, Erlangen, Germany
[2]Department of Physics, Friedrich-Alexander-University Erlangen-Nuremberg, Germany

*Corresponding author. Email: pascal.delhaye@mpl.mpg.de

**Abstract:** Advances in high-resolution laser spectroscopy have enabled many scientific breakthroughs in physics, chemistry, biology and astronomy. Optical frequency combs have pushed measurement limits with ultrahigh-frequency accuracy and fast-measurement speed while tunable diode laser spectroscopy is used in scenarios that require high power and continuous spectral coverage. Despite these advantages of tunable diode laser spectroscopy, it is challenging to precisely determine the instantaneous frequency of the laser because of fluctuations in the scan speed. Here we demonstrate a simple spectroscopy scheme with a frequency modulated diode laser that references the diode laser on-the-fly to a fiber cavity with sub-15 Hz frequency precision over an 11-THz range at a measurement speed of 1 THz/s. This is an improvement of more than two orders of magnitude compared to existing diode laser spectroscopy methods. Our scheme provides precise frequency calibration markers while simultaneously tracking the instantaneous scan speed of the laser. We demonstrate several applications, including dispersion measurement of an ultra-high-Q microresonator and spectroscopy of an HF gas cell, which can be used for absolute frequency referencing of the tunable diode laser. The simplicity, robustness and low costs of this spectroscopy scheme could prove extremely valuable for out-of-the-lab applications like LIDAR, gas spectroscopy and environmental monitoring.

Since the first demonstration of the laser in the 1960s, high-resolution laser spectroscopy has been an ultimate tool for studies on detailed structures and dynamics in atoms and molecules (*1*), further boosted by the progress of tunable continuous-wave (CW) lasers and optical frequency combs (*2–4*). In particular, the development of optical frequency combs enabled frequency metrology with unprecedented precision of up to 18 digits (*5*), leading to a rapidly growing number of applications (*6*, *7*). In addition to "single-frequency" metrology, optical frequency combs have also been utilized for high-precision and high-speed broadband spectroscopy, benefitting from their unique combination of large bandwidth and high spectral resolution. Within the last two decades, a diverse set of spectroscopic methods based on optical frequency combs have been developed, for example, direct frequency comb spectroscopy (*8*), dual-comb spectroscopy (*9*, *10*), two-dimensional optical frequency 'brushes' (*11*), Fourier transform spectroscopy (*12*), and a Vernier spectrometer (*13*). A disadvantage of spectroscopy with frequency combs is the low power per comb mode, which makes sensing of trace gas samples challenging. In addition, most combs exhibit a significant variation of power in different spectral regions, which can hinder broadband spectroscopy (*7*). As a further complication, high-precision measurements based on optical frequency combs often require comb sources with long-term coherence, which requires sophisticated servo loops (*6*).

Tunable CW lasers are widely used for high-sensitivity molecular spectroscopy and sensing with a high signal-to-noise ratio (SNR), because of their high photon flux, long interaction path, and frequency agility. However, the frequency resolution is limited by fluctuations of laser frequency scan speed. To address some of these issues, frequency-comb-calibrated diode laser spectroscopy was demonstrated by combining the accuracy of a frequency comb with the tunability and high power of a tunable laser, enabling broadband spectroscopy with bandwidth >4 THz, scanning speeds >1 THz/s, and MHz-level resolution (*14*). This method has wide applications in different areas, such as dispersion characterization of microresonators (*14–16*), absolute distance measurements (*17*, *18*), precision characterization of dynamic CW lasers (*19*, *20*), three-dimensional imaging (*21*), molecular spectroscopy (*22*), and calibration of astrophysical spectra (*23*). However, to generate frequency calibration markers with a high SNR, this method requires a reference frequency comb with a flat optical spectrum and constant polarization over a wide spectrum. So far, comb-calibrated diode laser spectroscopy has enabled tens of kilohertz frequency resolution (*20*, *24*). However, it is in particular challenging to measure very narrow spectral features across a broad spectral range with high power and high SNR. In addition, most frequency-comb-based applications are demonstrated in laboratory settings due to the requirement for a stable environment.

Here, we present a simple and easily accessible broadband spectroscopy scheme with Hz-level resolution based on a tunable diode laser. The diode laser frequency is referenced on-the-fly by a high finesse fiber cavity. The inclined reader might notice that this referencing to a fiber cavity poses two problems: (1), the resonances of a fiber cavity are not equally spaced due to dispersion, and (2), the rate at which the tunable laser changes its frequency suffers from fluctuations, which makes it difficult to predict its frequency into the future. We solve both of these problems by in-situ calibration of the high finesse fiber cavity mode spacing using sidebands modulated onto the tunable laser. As shown in Fig.1A, the resonances of a high-precision-calibrated dispersive cavity provides a series of frequency markers in time domain that are used to precisely determine the instantaneous frequency of a CW laser while tuning its wavelength. A dual radio frequency (RF) modulation scheme is used to calibrate the dispersive cavity on-the-fly with high precision, as shown in Fig.1B. Dual RF signals modulate the frequency of the tunable CW laser, which generates optical sidebands that are used to determine the cavity free spectral range (FSR). We apply this method to measure mode spectra of a fiber loop cavity with sub-15-Hz frequency resolution across an 11-THz spectral range within 10 seconds, revealing its complex dispersion profile. We use the calibrated cavity for several applications including gas absorption spectroscopy and the measurement of mode spectra of integrated photonic devices. This spectroscopy method enables a broad spectral range (limited by that of the tunable laser), fast measurement speed, ultrahigh frequency resolution, and high optical power. The relatively simple setup makes this method very robust for out-of-the-lab applications. We believe it can become a powerful tool for high-precision broadband spectroscopy, LIDAR, environmental monitoring, and characterization of photonic devices.

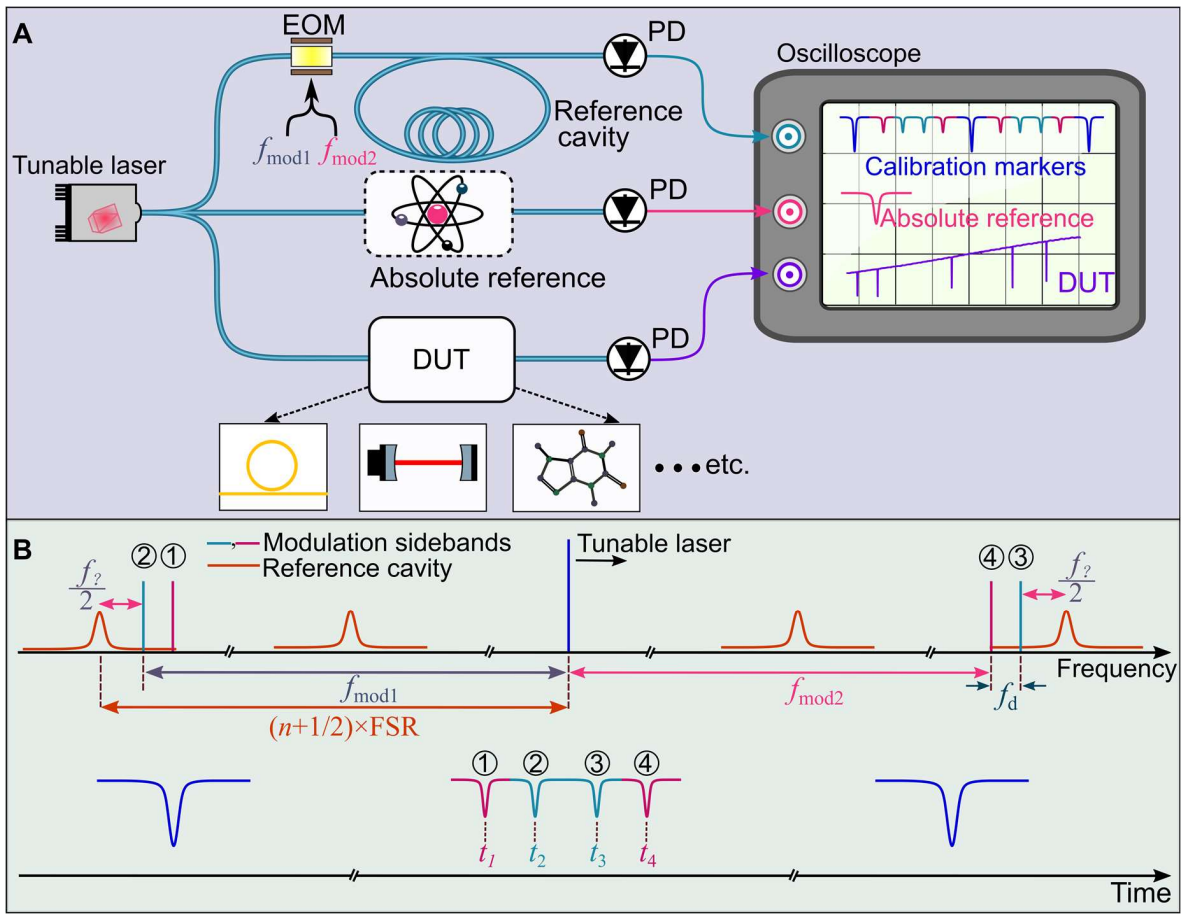

**Fig. 1. Principle of a Hz-level broadband spectrometer based on dual RF modulation.** (**A**) Measurement scheme. A tunable CW laser is modulated by two RF signals ($f_{mod1}$, $f_{mod2}$) via an electro-optic modulator (EOM). The modulated light is used to probe a reference cavity with quasi-periodic structures, such as a fiber cavity or integrated photonic cavity. The transmitted light is monitored by a photodiode (PD) and recorded by an oscilloscope to provide frequency reference markers for the scanning diode laser. The referenced diode laser is used to spectrally measure devices under test (DUT) such as on-chip photonic devices or gas absorption spectra. Optionally, part of the light probes a narrow linewidth atomic/molecular transition for an absolute frequency reference. (**B**) Principle of frequency to time conversion based on dual RF modulation. Dual RF modulation generates 2 sidebands neighboring the carrier frequency on both sides plotted on the frequency axis. While scanning the laser frequency, the generated sidebands form 4 additional transmission dips together with the carrier transmission dips in the time domain. The laser scan speed, laser frequency, and FSR can be precisely traced back to the modulation frequencies and their difference, and the time interval between the small sideband dips.

## Results

### Schematic and principle of a Hz-level broadband spectrometer

Figure 1 illustrates the schematic and the working principle of a Hz-level broadband spectrometer based on dual RF modulation. As shown in Fig. 1A, an external cavity diode laser with a widely tunable frequency range is modulated by two RF signals via an electro-optic intensity modulator (EOM), generating four optical sidebands in the frequency domain. The dual-RF modulated light is used to probe a reference cavity with an FSR smaller than or similar to the modulation frequency, such as fiber loop cavities, fiber linear cavities, Fabry-Perot cavities, etc. Important is only a high finesse and narrow linewidth of the reference cavity. Note that this reference cavity also exhibits dispersion, such that the FSR changes with wavelength. Depending on the FSR of the reference cavity, the frequency of the two RF signals $f_{mod1}$, $f_{mod2}$ ($f_{mod1} > f_{mod2}$) can be set to around $(n+1/2)\times$FSR, where $n$ is an integer number or zero. The frequency difference $f_d$ between the two RF signals can be set to around a few MHz, depending on the optical linewidth of the dispersive cavity. The transmitted or reflected light of the reference cavity is monitored by a photodiode and recorded by an oscilloscope. Figure 1B shows the principle of time to frequency conversion based on the dual RF modulation. As plotted in the frequency domain (upper part of Fig. 1B), we determine the moments in time when the laser carrier frequency $f_{carrier}$ is exactly in the middle between two reference cavity resonances, which are separated from each other by $(2n+1)\times$FSR. Just a tiny amount of time before the tunable laser is exactly in between two reference cavity resonances, the modulation sidebands (1) and (2) crossed a cavity resonance, generating calibration markers (1) and (2) in the time domain. When the tunable laser continues sweeping towards higher frequencies, the modulation sidebands (3) and (4) will generate to additional calibration markers. As a result, we measure four calibration marker transmission dips that are symmetric around the point in time when the tunable laser is exactly between two reference cavity resonances. These are used to determine the local FSR of the reference cavity with high accuracy as well as the scan speed of the laser. In addition to the 4 calibration markers is time domain, we also measure additional transmission dips when the tunable laser itself crosses the cavity resonances as shown in the lower part of Fig. 1B. This pattern repeats every time the tunable laser crosses one FSR of the reference cavity. As shown in Fig. 1B, the local FSR of the dispersive cavity is determined by $(2n + 1)\times$FSR $= 2f_{mod1} + f_?$, where $f_?/2$ is the frequency difference between the optical sideband $f_{mod1}$ and nearby cavity resonance when the carrier frequency lies in the middle of two cavity resonances. Since we know $f_{mod1}$ and $n$, we just need to determine $f_?$. From Fig. 1B, we can see that the laser took the time interval $(t_3-t_2)$ in order to scan across $f_?$. In addition, we know a precise value for the current laser scan speed, which is given by $v_{scan} = f_d/(t_2 - t_1)$ or $v_{scan} = f_d/(t_4 - t_3)$, where $f_d = f_{mod1} - f_{mod2}$ is the difference between the RF modulation frequencies. Thus, we obtain $f_? = v_{scan}\times(t_3 - t_2)$. We can take the average of the two ways to calculate the scan speed $v_{scan}$ in order to increase our measurement precision. Note that we can set the RF modulation frequencies such that all the calibration markers 1…4 are very close to each other in time, such that we obtain a nearly instantaneous calibration of the cavity FSR that is largely insensitive to fluctuations of the laser scan speed, which we assume to be constant within the tiny time interval $(t_3-t_2)$. Note also, that we need two RF modulation frequencies that are close to each other in order to determine an accurate value for the current laser scan speed. In comparison, for a single RF modulation scheme, the laser scanning speed would need to be calculated based on the average scanning speed across a whole FSR. This will induce relatively large frequency uncertainty in $f_?$. As shown in Fig. 1A, with the calibrated dispersive cavity and optionally together with an absolute frequency reference from atomic or molecular absorption lines, we can apply this spectroscopy method to measuring integrated photonic devices like microresonators, for open-path gas sensing, or generally optical frequency metrology. Important for out-of-lab applications, this scheme is very robust and does not require mode locked frequency combs or optical stabilization schemes, which could be useful in rough environments.

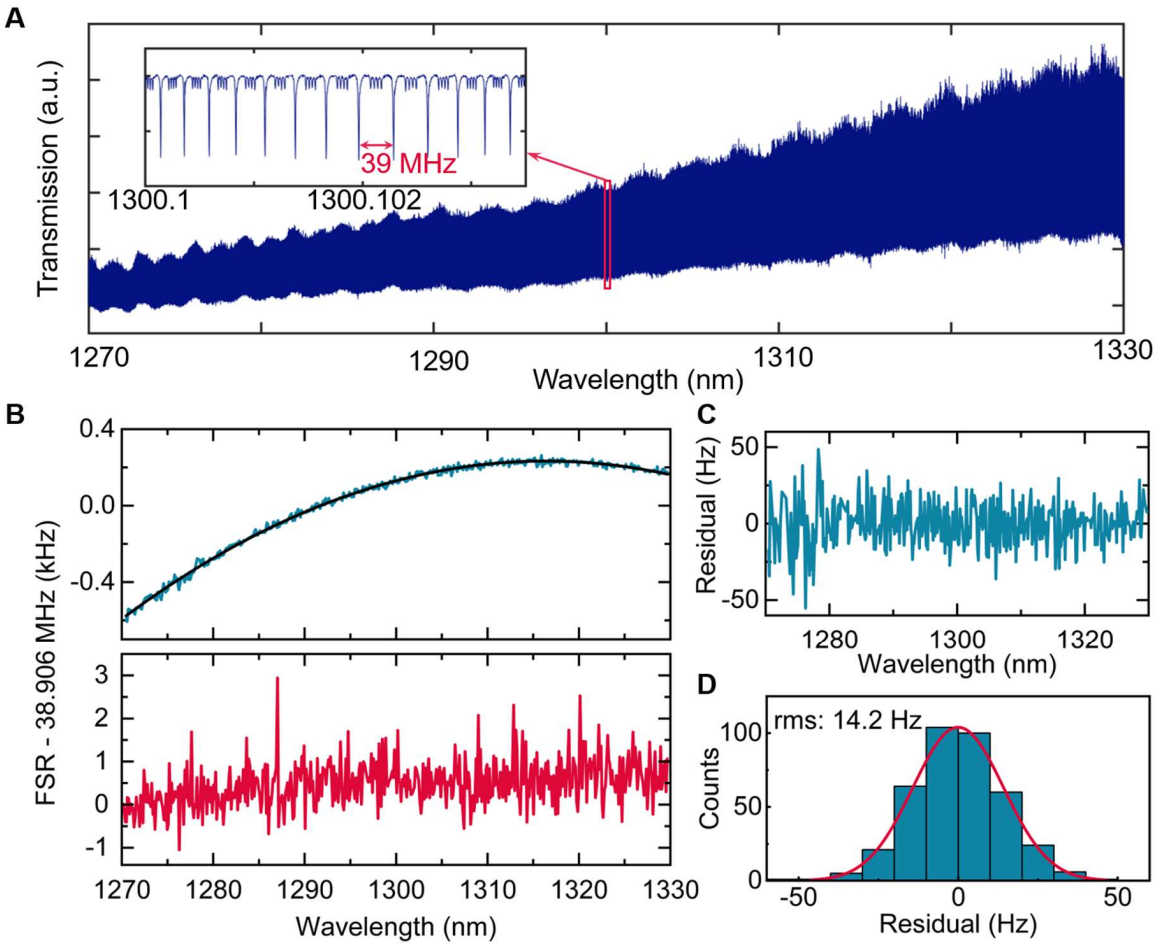

**Fig. 2. Mode spectrum measurement of a fiber loop cavity calibrated by dual RF modulation.** (**A**) Transmission spectrum of the 5-m fiber loop cavity. Inset: zoomed-in section showing the calibration markers around 1300 nm. The deep transmission dips are cavity resonances measured by the sweeping carrier laser while the small four dips within one FSR result from the RF modulation sidebands. (**B**) Measured FSR evolution (blue, left axis) of the fiber loop cavity interrogated by the dual RF modulation scheme, together with a second-order polynomial fit (black), in contrast to the result (red, right axis) measured by single RF modulation. (**C**) Frequency difference between the measured FSR and the fitted curve in (**B**). (**D**) Histogram of the data in (**C**) and a fitted Gaussian curve with a root-mean-square (rms) deviation of 14 Hz.

## Characterizing of a fiber loop cavity

As a proof-of-concept demonstration, we verify our scheme by measuring the dispersion of a fiber loop cavity. The fiber cavity is made from a 10-dB fiber coupler and 5 m standard telecom fiber (SMF-28) with a zero-dispersion wavelength of 1310 nm. The FSR of the fiber cavity is 39 MHz, and the mode linewidth is 1 MHz, which limits the measurement speed to around 1 THz/s (*14*). To demonstrate the ultrahigh frequency resolution of our scheme, we use a 1.3-µm laser with a tuning range from 1270 to 1330 nm in order to resolve the small FSR variation of the fiber cavity around the zero-dispersion wavelength. In the experiment, the tunable laser is modulated by two 20-GHz signals with a 4-MHz frequency difference. The two RF signals are combined with a power combiner and applied to the EOM. The light transmitted through the fiber cavity is detected by a photodiode (PD) and recorded using an oscilloscope with a memory depth of 31.25 million data points. Figure 2A shows the transmission spectrum of the fiber cavity from 1270 to 1330 nm and the inset shows an enlarged section of the data around 1300 nm. The deep transmission dips are fiber cavity resonances seen directly by the carrier, while the four additional small dips within one FSR result from the modulation sidebands crossing resonances. Figure 2B shows the measured FSR evolution of the fiber loop cavity as a function of wavelength around the zero-dispersion regime. The blue trace (upper panel) in Fig. 2B shows the results calculated from the dual RF modulation scheme. With a sub-15-Hz frequency resolution, the blue trace clearly resolves the small FSR variation (< 800 Hz) for a range of 11 THz and reveals the evolution of the cavity dispersion from normal at short wavelengths (FSR increasing with wavelength), crossing zero (region with nearly constant FSR), and to anomalous dispersion at longer wavelengths (FSR decreasing with wavelength). The black trace is the second-order polynomial fit with the zero-dispersion wavelength at 1315 nm. For comparison, the red trace (lower panel) in Fig. 2B shows the calculated FSR evolution based on a single RF modulation sideband (20 GHz modulation frequency), which cannot resolve the small changes of the cavity FSR. This shows that the dual RF modulation that allows to precisely determine the laser scan speed is critical for this spectroscopy method. Figure 2C shows the frequency difference of the measured FSRs with respect to the fitted values. Figure 2D shows a histogram of the frequency difference based on the data in Fig. 2c, with a root-mean-square deviation of 14.2 Hz. These results confirm the ultrahigh-frequency resolution of our dual RF broadband spectroscopy.

Figure 3A shows the calculated group velocity dispersion $\beta_2$ of the 5-m-long fiber cavity, based on the fitted data in Fig. 2B. The results agree well with the dispersion of standard telecom fiber. The blue trace in Fig. 3B shows the corresponding group delay dispersion (GDD) of the 5-m-long fiber loop cavity, which includes the dispersion of a 10-dB coupler that is part of the cavity. We can use this setup to precisely measure dispersion of optical fibers by adding fiber to the cavity and measuring the change in dispersion. The red trace in Fig. 3B shows data from another measurement with 3 m fiber removed from the cavity, leading to a 2-m-long cavity with a zero-dispersion wavelength of 1318 nm. By subtracting the GDD of the 2-m-long fiber cavity from that of the 5-m-long cavity, we can get the GDD of the removed 3-m-fiber, which is shown as black trace in Fig. 3B with a zero-dispersion wavelength of 1312 nm. From the measurements, we can see that the zero-dispersion wavelength of longer cavities approaches that of the used fiber.

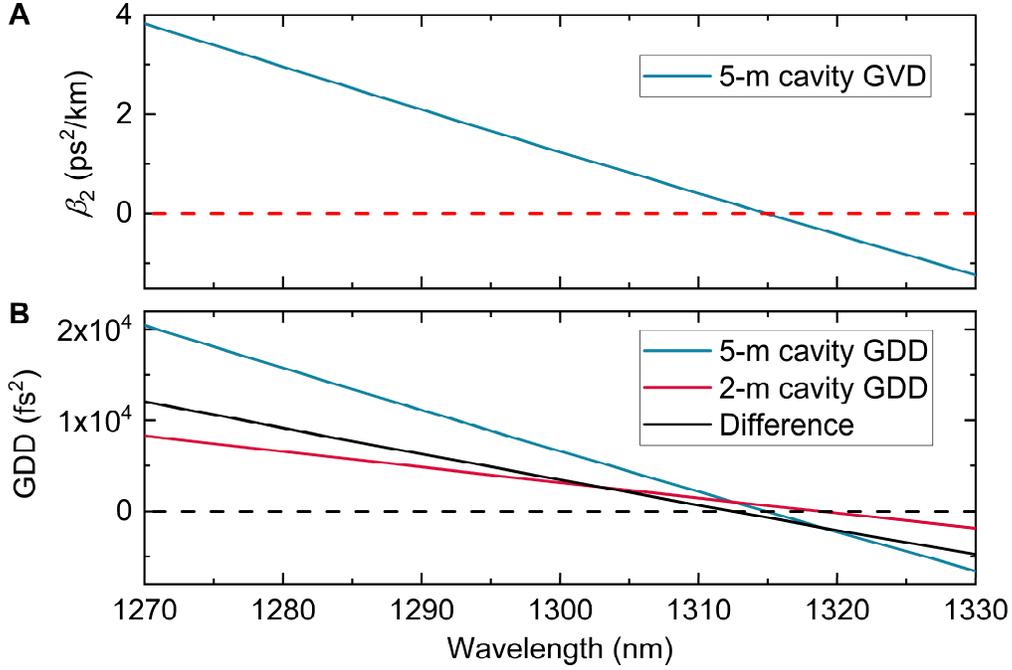

**Fig. 3. Measured dispersion of the fiber cavity** based on the fitted trace in Fig. 2B. (**A**) Group velocity dispersion of a 5-m fiber cavity. (**B**) Group delay dispersion of a 5-m-long fiber cavity (blue), 2-meter-long fiber cavity (red), and 3 meters of optical fiber.

## Characterizing an on-chip optical microresonator

Instead of measuring dispersion of optical fibers, our method can be easily applied to spectrally characterize the optical properties of different optical components, such as dispersion-engineered broadband mirrors, and integrated photonic circuits. We show this by characterizing the transmission spectrum of an optical microresonator (*25*). As shown in Fig. 1A, one part of the tunable CW laser is modulated by two RF frequencies and injected into the fiber cavity to generate precise reference markers that are recorded with an oscilloscope. This allows us to know the precise frequency of the laser at any given time during the sweep. We can now use another part of the laser for spectroscopy by simultaneously measuring a test device and recording the transmission signal on another channel of an oscilloscope, as shown in Fig. 1A. To demonstrate this technique, we use the resonance frequencies of the calibrated 5-m fiber cavity as frequency markers to measure the transmission spectrum of an in-house fabricated $Si_3N_4$ ring resonator. The $Si_3N_4$ resonator is made from a 750-nm $Si_3N_4$ thin film that is deposited onto a 3-μm $SiO_2$ intermediate layer on a silicon substrate via low-temperature reactive sputtering (*26*, *27*). The $Si_3N_4$ microresonator used in the experiments has a diameter of 200-μm and a waveguide cross-section of 1.8 μm × 750 nm. The resonator has an FSR of 231 GHz and an intrinsic optical quality factor of 3 million. The resonance frequencies of a mode family in a dispersive resonator can be described by a Taylor series as (*28*, *29*)

$$\begin{aligned}\omega_\mu &= \omega_0 + D_1\mu + \frac{D_2}{2!}\mu^2 + \frac{D_3}{3!}\mu^3 + \frac{D_4}{4!}\mu^4 + \ldots, \\ &= \omega_0 + D_1\mu + D_{\text{int}}(\mu)\end{aligned} \quad (1)$$

where $\mu$ is the mode number offset from the center mode at $\mu = 0$ and $\omega_\mu$ are the resonance frequencies. $D_1/2\pi$ is the FSR of the resonator at an arbitrarily chosen central mode ($\mu = 0$), and $D_2$, $D_3$, and $D_4$ are coefficients of second-, third-, and fourth-order dispersion, respectively. $D_{int}$ is the integrated dispersion, depicting the deviation of resonance frequencies from the equidistant grid spaced by $D_1$. While scanning the frequency of the CW laser, both transmission signals from the fiber cavity and $Si_3N_4$ resonator are simultaneously recorded. Figure 4A shows the normalized transmission spectrum of the $Si_3N_4$ resonator. Two different mode families are observed. The red markers highlight the fundamental TE mode family with a higher optical quality factor. Figure 4B shows one microresonator resonance (blue) around 1270.6 nm together with frequency markers (red) from the fiber cavity. The inset in Fig. 4B shows a scanning electron microscope image of the resonator.

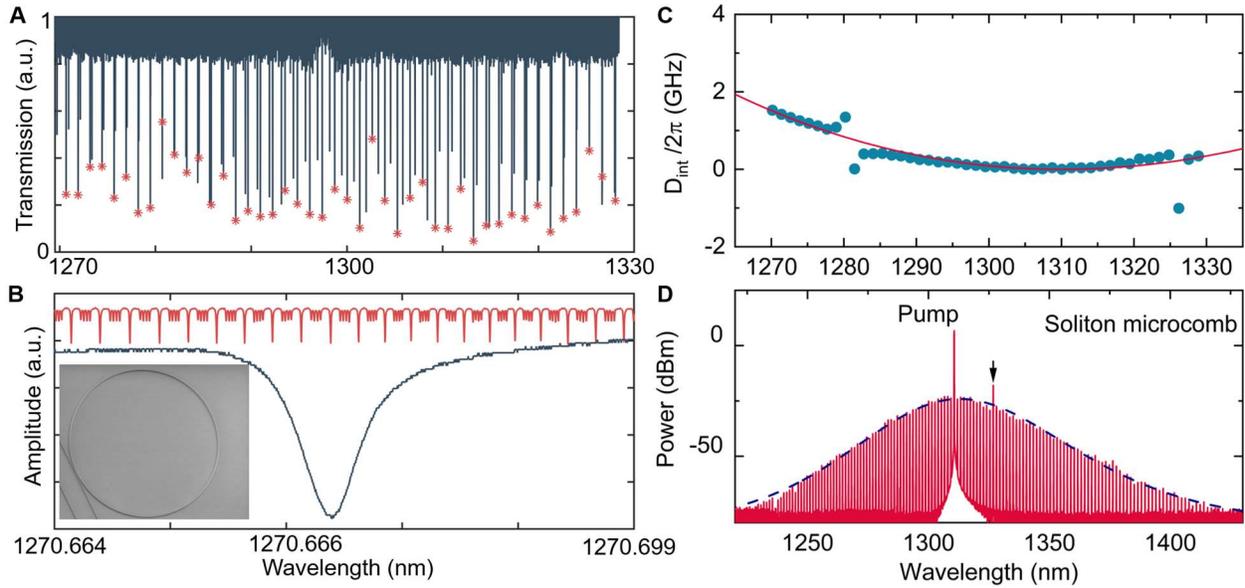

**Fig. 4. $Si_3N_4$ resonator mode spectrum and dispersion measurement.** (**A**) Normalized transmission spectrum of the $Si_3N_4$ on-chip resonator with red markers on the mode family with higher optical quality factor. (**B**) zoomed-in spectrum around 1270.6 nm together with fiber cavity resonance markers. Inset: scanning electron microscope image of the 200-µm-diameter $Si_3N_4$ resonator used in the experiments. (**C**) Measured integrated dispersion profile (blue circles) at pump wavelength of 1310 nm together with a second-order polynomial fit (red trace). (**D**) Optical spectrum of a bright soliton generated in the $Si_3N_4$ resonator pumped at 1310 nm and a $sech^2$ envelope fit (blue dashed trace). A dispersive wave (marked with an arrow) is observed at 1326 nm.

The in-situ measured FSR of the fiber cavity and the corresponding reference markers are used to convert the time axis of the oscilloscope into optical frequencies. This allows us to measure the mode structure of the $Si_3N_4$ resonator, including the evolution of its mode spacing, resonance linewidth, and dispersion. Figure 4C shows the measured integrated dispersion profile (blue circles) at a central mode at 1310 nm together with a second-order polynomial fit (red), depicting the deviation of resonance frequencies from the equidistant grid spaced by the FSR at the central mode, based on the resonances with red markers in Fig. 4A. The dispersion profile exhibits anomalous dispersion with mode crossings at 1280 and 1326 nm (*30*). To verify the dispersion measurement, we test this $Si_3N_4$ resonator for the generation of a soliton frequency comb by pumping an optical mode at 1310 nm (*31*, *32*). Figure 4D shows the optical spectrum of a single bright soliton with a fitted $sech^2$ envelope (blue dashed line) (*32*, *33*), which further verifies the anomalous dispersion regime for the pump mode. In addition, the mode crossing at 1326 nm in Fig. 4C induces a dispersive wave (marked with an arrow) at 1326 nm in Fig. 4D (*28*).

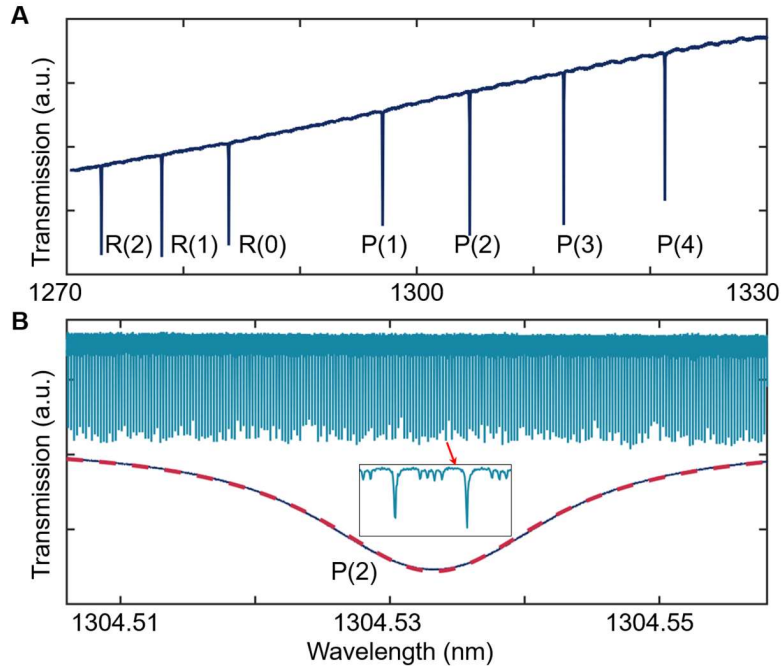

**Fig. 5. Absorption spectrum of a gas cell filled with molecular HF.** (**A**) Transmission spectrum of the HF gas cell from 1270 to 1330 nm. (**B**) Spectral profile (solid black line) of the HF P(2) line together with the frequency markers (blue) from the 5-m fiber cavity. The dashed red line is a Lorentzian fit.

**Hydrogen fluoride gas spectroscopy**

To highlight another application of this spectroscopy method, we demonstrate the measurement of non-periodic spectra, such as the absorption spectrum of a gas. A fiber coupled hydrogen fluoride ($H^{19}F$) gas cell with 50-Torr pressure and 2.7-cm optical path length is used for this demonstration. Just as before, the output light of the sweeping 1.3-µm CW laser is split into two paths, one part is referenced to the 5-m fiber cavity to generate calibration markers, while the other part is used for simultaneously probing the absorption spectrum of the gas cell. Figure 5A shows the strong HF molecular absorption lines (P and R branches) in the O-band range. Figure 5B shows the zoomed-in spectrum of the P(2) absorption line (solid black line) with fiber resonances (blue line) as frequency references. Since the pressure-broadening effect of the HF gas is much larger than its Doppler-broadening effect (*34*, *35*), a Lorentzian function (dashed red line in Fig. 5B) is used to fit the spectral profile. Table 1 shows the measured position of the HF absorption lines in comparison to the HITRAN database (*36*). Column 2 is the absorption line position calculated from the HITRAN database, corrected with the pressure shift from the vacuum transition wavelength. Columns 3 and 4 are the theoretically calculated Gaussian and Lorentzian full width at half-maximum (FWHM) linewidths, respectively. The uncertainty is calculated based on 20% uncertainty of the gas pressure, specified by the gas cell manufacturer. Column 5 shows the measured line positions. Here, we use the measured R(2) line as absolute frequency reference and set its frequency equal to the value calculated from HITRAN database. Column 6 is the wavelength difference between the measured results and the HITRAN database. The small discrepancies are within the error margin expected from the uncertainty of the gas pressure. Using the calculated Gaussian linewidth in column 3 and a Voigt fitting function, the last column shows the measured Lorentzian FWHM of the absorption lines. Considering that the positive pressure shift coefficient (cm$^{-1}$/atm) and the wavelength of the measured line positions is smaller than the calculated number in column 2, as well as that the measured Lorentzian (pressure broadened) linewidth is larger than the calculated number, we speculate that the pressure of the HF gas cell used in the experiments is slightly higher than the 50 Torr specified by the manufacturer.

**Table 1. Measured wavelengths and linewidth of HF absorption lines and comparison with values from the HITRAN database.**

| Line | HITRAN[1,2] (nm) | Cal. Gauss. $\Delta f$ (MHz) | Cal. Lorentz.[2] $\Delta f$ (MHz) | Measured position[3] (nm) | $\Delta$(pm) | Measured Lorentz. $\Delta f$ (MHz) |
|---|---|---|---|---|---|---|
| R(2) | 1272.97025 (± 0.04 pm) | 648.8 | 2453.6 (± 491.1) | 1272.97025 | - | 3704.1 |
| R(1) | 1278.14783 (± 0.04 pm) | 646.2 | 2496.9 (± 499.7) | 1278.14748 | -0.35 | 3756.2 |
| R(0) | 1283.88526 (± 0.09 pm) | 643.3 | 2260.3 (± 452.4) | 1283.88450 | -0.76 | 2909.6 |
| P(1) | 1297.07013 (± 0.03 pm) | 636.8 | 2469.3 (± 494.2) | 1297.06900 | -1.12 | 3008.6 |
| P(2) | 1304.53367 (± 0.04 pm) | 633.1 | 2741.5 (± 548.7) | 1304.53276 | -0.91 | 3500.8 |
| P(3) | 1312.59095 (± 0.02 pm) | 629.2 | 2650.8 (± 530.5) | 1312.59047 | -0.48 | 3168.9 |
| P(4) | 1321.25259 (± 0.02 pm) | 625.1 | 2039.4 (± 407.9) | 1321.25205 | -0.54 | 2259.3 |

1. Data from HITRAN are given after adding a 50-Torr pressure shift
2. The uncertainties of the pressure shift and pressure-broadened linewidths are calculated from the pressure uncertainty of 20%
3. The measured R(2) wavelength is set equal to the value from HITRAN and used as absolute reference

**Conclusion**

In summary, we have proposed and demonstrated a powerful broadband and high-precision spectroscopy technique based on a tunable CW laser. The laser frequency is on-the-fly calibrated using a fiber cavity and a dual RF frequency modulation technique that precisely determines the laser scan speed and provides calibration markers. By combining the frequency accuracy of radio frequency modulation sidebands with the short-term stability of a high finesse fiber cavity, we realize a broadband and easy-to-use optical frequency reference. Applying this method, we can measure miniscule FSR deviations of a fiber loop cavity at close-to-zero dispersion over an 11-THz frequency range with sub-15-Hz accuracy, which is an improvement of more than two orders of magnitude in comparison with existing comb-calibrated tunable laser spectroscopy methods. The demonstrated measurement speed is 1 THz/s, which was limited by the cavity linewidth of the tested microresonator mode and corresponding cavity build-up time. Compared to frequency comb-based spectroscopy, this scheme provides high optical probe power as well as better spectral flatness and polarization stability. In addition, this method could enable access to new wavelength regions, e.g. in the mid infrared or UV. We demonstrate several applications by characterizing the dispersion of an integrated photonic microresonator device and measuring the molecular absorption spectrum of HF gas. By using a well-known atomic/molecular transition as an absolute frequency reference, this method can be used for highly accurate broadband molecular spectroscopy (*37*), providing a microwave link between optical frequencies. In addition, without the requirement of mode locking or phase locking, this simple and robust method can find applications in out-of-lab applications, such as LIDAR systems (*17*), 3D imaging (*21*), refractive index measurements (*38*), industrial and agricultural monitoring (*39–41*), characterization of photonic devices (*28*, *42*), and calibration of astrophysical spectrometers (*43*).

**Funding:** This work is supported by European Union's H2020 ERC Starting Grant "CounterLight" 756966; H2020 Marie Sklodowska-Curie COFUND "Multiply" 713694; Marie Curie Innovative Training Network "Microcombs" 812818; and the Max Planck Society.

**Author contributions:** S.Z. conceived and performed the experiments. S.Z., T.B., and P.D. analyzed the data. All co-authors contributed to the manuscript.

**Competing interests:** The authors filed a patent application for the demonstrated spectroscopy method.

**Data and materials availability:** The data that support the findings of this study are available from the corresponding authors upon reasonable request.